\documentstyle[12pt]{article}

\catcode`@=11 \@addtoreset{equation}{section} \catcode`@=12

\newcommand\remove[1]{}
\newcommand{\f}{\frac}
\newcommand{\p}{\partial}
\newcommand{\T}{\tilde}
\newcommand{\s}{\sqrt}
\newcommand{\del}{\delta}
\newcommand{\ep}{\epsilon}
\newcommand{\varep}{\varepsilon}
\newcommand{\g}{\gamma}
\newcommand{\G}{\Gamma}
\newcommand{\k}{\kappa}
\newcommand{\om}{\omega}
\newcommand\e{{\rm e}}
\begin{document}
\begin{titlepage}
\begin{flushright}
IP/BBSR/2002-03\\
hep-th/0206032
\end{flushright}

\vspace{.2in}
\begin{center}
{\Large \bf Curved Membrane Solutions in D=11 Supergravity}
\vspace{10mm}

{\large Sandip Bhattacharyya$^{*}$, Alok Kumar$^{*}$ and 
Subir Mukhopadhyay$^{**}$}\\
\vspace{8mm}
$^{*}${\em Institute of Physics, Bhubaneswar 751 005, India}\\
$^{**}${\em Department of Physics, University of Massachusetts,
Amherst, MA 01003}\\
email: sandip,kumar@iopb.res.in, subir@physics.umass.edu
\end{center}
\begin{abstract} 
\begin{quotation}\noindent
\baselineskip 10pt
We present a class of static membrane solutions, with non-flat worldvolume
geometry, in the eleven dimensional supergravity with source terms.
This class of solutions contains supersymmetric as well as a large class of 
non-supersymmetric configurations. We comment about near horizon limit
and stability of these solutions and point out an interesting relation
with certain two dimensional dilaton gravity system. 
\end{quotation}
\vspace {.2in}
\end{abstract}
\vfill
\eject
\end{titlepage}

\section{Introduction}
\label{Intro} 

Branes with curved geometries\cite{papa} are objects of great
interest in finding nonperturbative dynamics in general backgrounds.  
Spherical and other geometries in this context have been a subject of 
interest for long\cite{kabat}. 
In particular, D-branes in curved 
backgrounds\cite{douglas} and their dynamics have been studied from 
the point of view of conformal field theory\cite{wzw}, as well as 
using the Born-Infeld action in curved geometries such as WZW 
models\cite{douglas-bachas}. Inspired by these developments, we 
study in this paper the eleven dimensional supergravity, 
understood to be the low  energy limit of M-theory, and analyze 
the possibility of obtaining a large class of non-flat membrane solutions. 

\remove {}
The existence of a supergravity in D=11 is known for a long time\cite{julia}.
The massless spectrum contains a graviton, a gravitino and a three form 
gauge field. Since a string  couples to a two form, it is only natural to
expect a membrane to exist in the 11 dimensional theory which will couple
to the three form. Indeed, such a membrane was eventually constructed
that couples to this background\cite{sezgin}. Subsequently, it emerged
as an exact solution of the supergravity field equations\cite{duff}. This 
solution has several interesting features. Firstly, the membrane has
an electric charge that is conserved due to the equations of motion.
Also, it has a $\del$-function singularity so that it cannot be
called a soliton in the usual parlance. Nevertheless, as is common with
the soliton solutions, they break just one half of the spacetime 
supersymmetry and saturate a Bogomol'nyi bound between the mass per 
unit area and the conserved charge. As a result, one can stack an 
arbitrary number of membranes together without affecting their
stability. After the connection between D=11 supergravity and the 
strong coupling limit of type IIA string theory was pointed out
\cite{witten}, it was observed that this membrane becomes the  
D2-brane of type IIA theory. Also, under double dimensional 
reduction, the membrane goes over to the well known macroscopic 
string solution of Dabholkar et al\cite{dabholkar}.

The flat membrane solution of \cite{duff} 
was constructed by assuming 
dependence of the classical background on coordinates transverse to 
the membrane. As a result, one has translational isometries for these 
solutions along two spatial directions in addition to a time translation, 
leading to Poincar\'{e} invariance. To find out a non-flat membrane 
solution, we use an ansatz which is a variation of the one in \cite{duff} 
by introducing dependence on spatial longitudinal coordinates. In particular, 
we now introduce metric components in the two spatial directions along the 
membrane, parametrized by a single conformal degree of freedom. In addition, 
we modify the radial dependence of the membrane solution by a factor dependent
on the world volume spatial coordinates. Finally, the 3-form ansatz is also 
modified by a similar factor.  

The plan of the paper is as follows. In section 2, we give 
a review of the membrane solution as obtained in \cite{duff}. 
In section 3, we present our solution. Unlike the flat membrane, 
our ansatz has nonvanishing curvature on the worldvolume. As a 
result, generically it breaks all supersymmetry. We then solve 
the supergravity equations of motion with a membrane source term. 
While solving the equations of motion, we discern an interesting 
connection of our solutions with the solutions of two dimensional 
dilaton gravity \cite{cadoni}. For a class of solutions of dilaton 
gravity, there exists a corresponding class of curved membrane 
solutions. We also find, somewhat to our surprise, that these
solutions have the same $AdS_4 \times S^7$ limit as that of a
flat membrane. In section 4, we conclude by discussing the stability 
of these solutions against small perturbations and point out some 
open problems. We give various explicit mathematical expressions in 
an appendix. 

\section{Flat membrane solution} 
For later convenience we shall review the flat membrane 
solution in this section.
Following \cite{duff}, one starts with an ansatz for the $D=11$ fields
$g_{MN}$ and $A_{MNP}$ ($M,N= 0,1,\ldots ,10$) corresponding to the
most general three-eight split invariant under $P_3 \times SO(8)$, where
$P_3$ is the three dimensional Poincar\'{e} group and $SO(8)$ is the eight
dimensional rotation group. The $D=11$ coordinates $x^M=(x^\mu,y^m)$,
where $\mu=0,1,2$ and $m=3,\ldots,10$.
The metric is:
\begin{equation}
ds^2 = \e^{2{\T{A}}}{\eta}_{\mu\nu}dx^{\mu}dx^{\nu} 
+\e^{2{\T{B}}}{\del}_{m n}dy^m dy^n,
\label{metric}
\end{equation}
and the three-form gauge field is:
\begin{equation}
A_{\mu\nu\rho} = \pm \f{1}{^{3}g} \varep_{\mu\nu\rho}\e^{\T{C}},
\label{3form}
\end{equation}
where $^{3}g$ is the determinant of $g_{\mu\nu}$ and 
$\varep_{\mu\nu\rho}$ is the three dimensional Levi-civita tensor.
All other components of $A_{MNP}$ and all components of the
gravitino $\psi_M$ are set to zero. Invariance under both $P_3$ and
$SO(8)$ forces the arbitrary functions $\T A,\T B,\T C$
to depend on $r \equiv \s{(y^m)^2}$.
The requirement of some unbroken supersymmetry relates $\T A$
and $\T B$ to $\T C$ so that we are left with only one 
undetermined function $\T C$. To preserve some supersymmetry,
there must exist Killing spinors $\varep$ satisfying 
\begin{equation}
{\T D_M}\varep = 0,
\label{kspeq0}
\end{equation} where
\begin{equation}
{\T D}_M = \p_M+\f{1}{4}{{\om}_M}^{AB}\G_{AB}
-\f{1}{288}({\G^{PQRS}}_M +8\G^{PQR}{{\del}^S}_M)
F_{PQRS}
\label{covder}
\end{equation}
with $F_{MNPQ}=4{\p_{[M}}A_{{NPQ}]}$.
Here ${{\om}_M}^{AB}$ are the spin connections and $\G_A$
are the $D=11$ Dirac matrices satisfying $\{\G_A,\G_B\} = 2\eta_{AB}$.
A,B refer to $D=11$ tangent space, $\eta_{AB}=diag(-,+,\ldots ,+)$
and $\G_{AB\ldots C}= \G_{[A}\G_B \ldots \G_{C]}$.
We then make a three-eight split:
$\G_A= ({\g}_{\alpha}\otimes\G_9, I \otimes{\Sigma}_a)$,
where $\g_\alpha$ and $\Sigma_a$ are the $D=3$ and $D=8$ Dirac 
matrices respectively and $\G_9=\Sigma_3 \ldots \Sigma_{10}$.

The most general spinor consistent with $P_3\times SO(8)$ is of the form
$\varep(x,y)=\ep \otimes \eta(r)$ where $\ep$ is a
constant spinor of $SO(1,2)$ and $\eta$ is an $SO(8)$ spinor.
After some calculation, we find that (\ref{kspeq0}) admits two non-trivial
solutions $(1\pm\G_9)\eta=0$ with $\eta=\e^{-\f{\T C}{6}}\eta_0$
and $\T A=\f{1}{3}\T C$, $\T B=-\f{1}{6}\T C$+ const.
In each case, one half of the maximal possible rigid supersymmetry survives. 
At this stage, the three unknown functions $\T A$, $\T B$ and
$\T C$ have been reduced to one by choosing the case where half 
supersymmetry survives. To determine this unknown function, we
substitute the ansatz into the field equations which follow from the action
\begin{equation}
S_G=\int d^{11} x {\cal L}_G,
\label{action}
\end{equation} 
where ${\cal L}_G$ is the supergravity lagrangian whose bosonic sector
is given by
\begin{eqnarray}
& &\k^2 {\cal L}_G =\f{1}{2}{\s{-g}}R-\f{1}{96}\s{-g}
F_{MNPQ}F^{MNPQ}+ \nonumber \\
& &\f{1}{2(12)^4}\varep^{MNOPQRSTUVW}F_{MNOP}F_{QRST}A_{UVW}.
\label{lagrangian}
\end{eqnarray}
The three form field equation is given by
\begin{equation}
\p_M(\s{-g}F^{MUVW})+\f{1}{1152}\varep^{UVWMNOPQRST}
F_{MNOP}F_{QRST} = 0.
\label{Feqn}
\end{equation}
The equation(\ref{Feqn}), with the above ansatz, leads to $\del^{mn}\p_m \p_n
\e^{-\T{C}} = 0$. Imposing the boundary condition that the metric
be asymptotically Minkowskian, we find $\e^{-\T {C}}=1+\f{K}{r^6}$,
$r > 0$ where $K$ is a constant.
Hence we get,
\begin{equation}
ds^2 = (1+\f{K}{r^6})^{-\f{2}{3}}\eta_{\mu\nu}dx^{\mu}dx^{\nu}
+(1+\f{K}{r^6})^{\f{1}{3}}\del_{mn}dy^m dy^n,
A_{\mu\nu\rho} = \pm
\f{1}{^{3}g}\varep_{\mu\nu\rho}(1+\f{K}{r^6})^{-1}.
\label{fldconfig}
\end{equation}
These expressions solve the field equations everywhere except at $r=0$.
Hence, to obtain a solution that is valid everywhere, we have to
modify the pure supergravity action by adding a membrane source
at $r=0$. Let us consider the combined supergravity and membrane 
equations which follow from the action
\begin{equation}
S=S_G+S_M,
\label{combaction} 
\end{equation}
where 
\begin{equation}
S_M = T\int d^3 \xi (-\f{1}{2}\s{-\g}\g^{ij}\p_i
X^M \p_j X^N g_{MN}+\f{1}{2}\s{-\g}+\f{1}{3!}
\varep^{ijk}\p_i X^M \p_j X^N \p_k X^P A_{MNP}),
\label{wvaction}
\end{equation}
where $T$ is the membrane tension. The Einstein equations are now
\begin{equation}
R_{MN}-\f{1}{2}g_{MN}R=\k^2 T_{MN},
\label{fldeqn}
\end{equation}
where,
\begin{eqnarray}
& &\k^2 T_{MN} = \f{1}{12}(F_{MPQR}F_N^{PQR}-\f{1}{8}g_{MN}
F_{PQRS}F^{PQRS})- \nonumber \\
& &\k^2 T \int d^3 \xi \s{-\g}\g^{ij}
\p_i X_M \p_j X_N \f{\del^{11}(x-X)}{\s{-g}},
\label{emtensor}
\end{eqnarray}
while the three form equation is:
\begin{eqnarray}
& &\p_M(\s{-g}F^{MUVW})+\f{1}{1152}\varep^{UVWMNOPQRST}
F_{MNOP}F_{QRST}= \nonumber \\
& &\pm 2\k^2 T \int d^3 \xi \varep^{ijk}\p_i X^U 
\p_j X^V \p_k X^W \del^{11}(x-X).
\label{Feqnfinal}
\end{eqnarray}
Also, we have the membrane field equations:
\begin{eqnarray}
& &\p_i(\s{-\g}\g^{ij}\p_j X^N g_{MN})+
\nonumber \\
& &\f{1}{2}\s{-\g}\g^{ij}\p_i X^N \p_j X^P
\p_M g_{NP}\pm \f{1}{3!}\varep^{ijk}\p_i X^N
\p_j X^P \p_k X^Q F_{MNPQ}=0,
\label{wveqn}
\end{eqnarray}
\begin{equation}
\g_{ij}=\p_i X^M \p_j X^N g_{MN}.
\label{pullback}
\end{equation}
One can easily verify that the correct source term is obtained by the
the static gauge choice $X^\mu =\xi^\mu$ and $Y^m$ =const, provided
$K=\f{\k^2 T}{3\Omega_7}$ where $\Omega_7$ is the volume of
$S^7$.
\section{A class of curved membrane solutions}
We now proceed to present our solution. Our ansatz is as follows:
\begin{eqnarray}
ds^2 = \e^{2A}[-(dx^0)^2+\s{f}(dx^a)^2]+\e^{2B}(dy^m)^2, 
\nonumber\\ \mbox{and}\ 
A_{\mu\nu\rho}=\pm\f{1}{^{3}g}\varep_{\mu\nu\rho}\e^C,
\nonumber\\ \mbox{with}\ 
\e^{2A} =\e^{2\T A}F^e , \e^{2B} =\e^{2\T B}g^b,
\e^C =\e^{\T C}\chi^c , r^6 =h^d{\T r}^6,\nonumber\\ 
\T r=\s{(y^m)^2}.
\label{nuansatz}
\end{eqnarray}
$\T A,\T B,\T C$ are functions of $r$ and $F,g,\chi,f,h$ are 
functions of $x^a$. In the above ansatz, $\T A,\T B,\T C$ are 
related to each other as in section 2, viz.,$\T A=\f{1}{3}\T C$, 
$\T B=-\f{1}{6}\T C$+ const. As is noticed, the worldvolume 
directions of the membrane for our ansatz is now curved due to the
introduction of the conformal factor $f$ above. In our ansatz, we have
chosen this factor along purely spatial directions $x^{1, 2}$. 
Alternatively, it is
possible to have a conformal factor along $x^0$ and $x^1$. We mainly 
restrict our discussion to the former possibility.

\subsection{Supersymmetry}
We would like to see whether the ansatz we propose preserves some amount
of space-time supersymmetry. For that we once again resort to the Killing
spinor equation.We give below only the necessary equations and refer the
reader to the appendix for more details.

We break the $SO(1,10)$ spinor $\varep$ into a $SO(1,2)$ spinor
$\ep$ and a $SO(8)$ spinor $\eta$. But now $\ep$ depends on
$x^{1,2}$ and $\eta$ depends on $r$. In our notation,
the three dimensional Dirac matrices are $\g_{\hat 0}=i\sigma_{2}$, 
$\g_{\hat 1}=\sigma_{1}$ and  $\g_{\hat 2}=\sigma_{3}$. The eight
dimensional Dirac matrices are denoted by $\Sigma_{m}$ as
usual. Also, the hatted indices denote tangent space indices
and the unhatted indices denote space-time indices. We now
obtain the Killing spinor equations for our ansatz using 
equation (\ref{kspeq0}). 

\begin{eqnarray}
{\T D}_0\varep & = & \f{d}{6}f^{-\f{1}{4}}(1+\f{K}{r^6})^{-1}\f{h_{,a}}{h}
(\g_{\hat 0}\g_{\hat a}\ep)\otimes \eta - \nonumber\\
               &   &\f{1}{6}(1+\f{K}{r^6})^{-\f{3}{2}}h^{-\f{d}{2}}
(\f{K}{r^6})_{,m}(\g_{\hat 0}\ep)\otimes (\Sigma_{\hat m}\G_9\eta)-
\nonumber\\
               &   &\f{1}{6}(1+\f{K}{r^6})^{-\f{3}{2}}h^{-\f{d}{2}}
(\f{K}{r^6})_{,m}(\g_{\hat 0}\ep)\otimes (\Sigma_{\hat m}\eta)=0,
\label{kspeq1}
\end{eqnarray}
\begin{eqnarray}
{\T D}_a\varep & = & \p_a\ep\otimes\eta+\f{d}{6}(1+\f{K}{r^6})^{-1}
\ep_{a\hat b}\f{h_{,\hat b}}{h}\g_0\ep\otimes\eta-\nonumber\\
               &   &\f{1}{4}\ep_{a\hat b}\f{f_{,\hat b}}{f}
\g_0\ep\otimes\eta+\nonumber\\
               &   &\f{1}{6}(1+\f{K}{r^6})^{-\f{3}{2}}h^{-\f{d}{2}}
f^{\f{1}{4}}(\f{K}{r^6})_{,m}(\g_{\hat a}\ep)\otimes(\Sigma_{\hat m}\G_9\eta)+
\nonumber\\
               &   &\f{1}{6}(1+\f{K}{r^6})^{-\f{3}{2}}h^{-\f{d}{2}}
f^{\f{1}{4}}(\f{K}{r^6})_{,m}(\g_{\hat a}\ep)\otimes(\Sigma_{\hat m}\eta)=0,
\label{kspeq2}
\end{eqnarray}
\begin{eqnarray}
{\T D}_m\varep & = & \ep\otimes\p_m\eta+\f{d}{12}(1+\f{K}{r^6})^{-\f{1}{2}}
h^{\f{d}{2}}f^{-\f{1}{4}}(\f{h_{,\hat a}}{h}\g_{\hat a}\ep)\otimes
(\Sigma_m\G_9\eta)\nonumber\\
               &   &+\f{1}{12}(1+\f{K}{r^6})^{-1}\ep\otimes
(\f{K}{r^6})_{,\hat n}\Sigma_{m\hat n}\eta \nonumber\\
               &   &+\f{1}{12}(1+\f{K}{r^6})^{-1}\ep\otimes
(\f{K}{r^6})_{,\hat n}\Sigma_{m\hat n}\G_9\eta \nonumber\\
               &   &+\f{1}{6}(1+\f{K}{r^6})^{-1}(\f{K}{r^6})_{,m}
\ep\otimes\G_9\eta=0.
\label{kspeq3} 
\end{eqnarray}
We will now discuss various cases.

1. \underline {1/2 SUSY solution:}

If we set $h_{,\hat a}=f_{,\hat a}=0$, we end up with a single condition
on $\eta$, viz. $(1+\G_9)\eta=0$ from (\ref{kspeq1}) and (\ref{kspeq2}). 
Finally, (\ref{kspeq3}) implies that $\eta=(1+\f{K}{r^6})^{\f{1}{6}}
\eta_0$, where $\eta_0$ is a constant spinor. As a result, the membrane 
breaks 1/2 SUSY. This is the well known Duff-Stelle solution\cite{duff}.

2. \underline{1/4 SUSY solution:}

We can have 1/4 SUSY solution for the following conditions:
\begin{equation}
h_{,\bar z}=f_{,\bar z}=0, \g_{z}\ep=0,
\label{susy}
\end{equation}
in addition to the condition for the 1/2 SUSY case.
Here the subscripts $z(\bar z)$ refer to the complex coordinates 
$x^1+ix^2$ and $x^1-ix^2$ respectively.

In fact, one can obtain other curved membrane solutions by interchanging
the role of one of the time coordinate with one of the worldvolume 
directions of the membrane. One then has a solution of the traveling
wave type. These statements apply to the nonsupersymmetric cases
discussed below as well. 

3. \underline{Nonsupersymmetric solutions:}

For generic choice of $h$ and $f$, supersymmetry is completely
broken. In the following, we discuss these solutions in detail.
 
\subsection{Field Equations}
The three form equation(\ref{Feqnfinal}) once again leads to
\begin{equation}
\e^{-\T {C}}=1+\f{K}{r^6},
\label{Ceqn}
\end{equation}
provided we impose the condition
\begin{equation}
F^{-\f{3}{2}e}g^{3b}\chi^c f^{-\f{1}{2}}h^{-d}=1.
\label{condsugra}
\end{equation} 
Similarly, the membrane field equation(\ref{wveqn}) gives the condition
\begin{equation}
F^{-{\f{3}{2}}e}f^{-\f{1}{2}}\chi^c  = 1.
\label{condmemb}
\end{equation}

Using (\ref{fldeqn}) and (\ref{emtensor}), we can write the Einstein
equations in the form
\begin{equation}
R_{MN} = \k^2(T_{MN}-\f{1}{9}g_{MN}T),
\label{RMNeqn}
\end{equation}
The equations $R_{0a}=0$ and $R_{0m}=0$ are identically satisfied by our 
ansatz. The equation $R_{am}=0$ is satisfied provided $g=h$ and $d=3b$. 
For the rest of the equations, we give below the details.
\begin{eqnarray}
& &R_{00} = \f{1}{3}F^{e}g^{-b}(1+\f{K}{r^6})^{-3}[(\f{K}{r^6})_{,m}]^2-
\f{1}{3}F^{e}g^{-b}(1+\f{K}{r^6})^{-2}(\f{K}{r^6})_{,mm}+\nonumber\\
& &\f{e}{2\s{f}}[\f{F_{,aa}}{F}+(\f{e}{2}-1)(\f{F_{,a}}{F})^2
+4b\f{g_{,a}}{g}\f{F_{,a}}{F}]+\nonumber\\
& &\f{d}{3}(\f{K}{r^6})(1+\f{K}{r^6})^{-1}\f{1}{\s{f}}
[\f{h_{,aa}}{h}-(d+1)(\f{h_{,a}}{h})^2-e\f{h_{,a}}{h}\f{F_{,a}}{F}
+4b\f{g_{,a}}{g}\f{h_{,a}}{h}], \nonumber\\
\cr
& &R_{ab} = d(\f{K}{r^6})(1+\f{K}{r^6})^{-1}[\f{h_{,ab}}{h}+
(d-1)\f{h_{,a}h_{,b}}{h^2}-
\f{1}{4hf}(f_{,a}h_{,b}+
f_{,b}h_{,a}-f_{,c}h_{,c}\del_{ab})]-\nonumber\\
& &\f{d}{3}(\f{K}{r^6})(1+\f{K}{r^6})^{-1}
[\f{h_{,cc}}{h}+(d-1)(\f{h_{,c}}{h})^2]\del_{ab}-
\f{1}{3}(g^{-b}F^{e}\s{f})(1+\f{K}{r^6})^{-3}
[(\f{K}{r^6})_{,m}]^2\del_{ab}+\nonumber\\
& &\f{1}{3}(g^{-b}F^{e}\s{f})(1+\f{K}{r^6})^{-2}
(\f{K}{r^6})_{,mm}\del_{ab},\nonumber\\
\cr
& &R_{mn} = -\f{b}{2}(F^{-e}g^{b-1}f^{-\f{1}{2}})
\del_{mn}[g_{,aa}+(4b-1)\f{{g_{,a}}^2}{g}+
\f{e}{2}\f{g_{,a}F_{,a}}{F}]-\nonumber\\
& &\f{1}{6}(1+\f{K}{r^6})^{-1}(\f{K}{r^6})_{,pp}\del_{mn}-
\f{1}{2}(1+\f{K}{r^6})^{-2}(\f{K}{r^6})_{,m}(\f{K}{r^6})_{,n}+\nonumber\\
& &\f{1}{6}(1+\f{K}{r^6})^{-2}[(\f{K}{r^6})_{,p}]^2\del_{mn}. 
\label{eom}
\end{eqnarray}
In writing $R_{ab}$ components in equation (\ref{eom}) we have
dropped terms independent of $r$. This can be done self-consistently
for our ansatz, as we discuss below after equation (\ref{2dgravred}),
using equation of motion (\ref{RMNeqn}).   
The source terms in the three cases are as follows.
\begin{eqnarray}
& &\k^2(T_{00} - \f{1}{9}g_{00}T)
= \f{1}{3}F^{e}g^{-b}(1+\f{K}{r^6})^{-3}[(\f{K}{r^6})_{,m}]^2+\nonumber\\
& &\f{2}{3}\k^2 T\int d^3\xi F^e g^{-4b}(1+\f{K}{r^6})^{-2}
\del^{11}(x-X),\nonumber\\
& &\k^2(T_{ab} - \f{1}{9}g_{ab}T)
= -\f{1}{3}(g^{-b}F^{e}\s{f})(1+\f{K}{r^6})^{-3}
[(\f{K}{r^6})_{,m}]^2\del_{ab}-\nonumber\\
& &\f{2}{3}\k^2 T \int d^3\xi(g^{-b}F^{e}\s{f})(1+\f{K}{r^6})^{-2}
g^{-d}\del^{11}(x-X),\nonumber\\
& &\k^2(T_{mn} - \f{1}{9}g_{mn}T)
= -\f{1}{2}(1+\f{K}{r^6})^{-2}(\f{K}{r^6})_{,m}(\f{K}{r^6})_{,n}+\nonumber\\
& &\f{1}{6}(1+\f{K}{r^6})^{-2}[(\f{K}{r^6})_{,p}]^2\del_{mn}+\nonumber\\
& &\f{1}{3}\k^2 T\int d^3\xi g^{-d}(1+\f{K}{r^6})^{-1}
\del^{11}(x-X).
\label{source}
\end{eqnarray}
The ansatz solves the equations of motion provided we set 
\begin{equation}
F=g=h,\ 2d= -3e = 6b. 
\label{relation}
\end{equation}
We also obtain $K=\f{\k^2 T}{3\Omega_7}$. But what actually makes 
the solutions interesting is the following pair of equations:
\begin{eqnarray}
R_{ab}(\s{f}F^e)+\f{3e}{2}\nabla_a\nabla_b\ln F-\f{3e^2}
{4}\nabla_a\ln F\nabla_b\ln F = 0,\nonumber\\
\nabla^2(\e^{-{\f{3e}{2}\ln F}}) = 0.
\label{2dgrav}
\end{eqnarray}
which arise from (\ref{RMNeqn}) using (\ref{eom})-(\ref{relation}).
In (\ref{2dgrav}), $R_{ab}(\s{f}F^e)$ denotes the Ricci tensor
components with conformal metric, $g_{ab}=\s{f}F^e\del_{ab}$.
More precisely, from the $(00)$ components in (\ref{eom}) and
(\ref{source}) we get the conditions, using(\ref{relation}):
\begin{eqnarray}
\f{F_{,ab}}{F} + (d-1)\f{F_{,a}F_{,b}}{F^2} - 
\f{1}{4}\f{1}{Ff}(f_{,a}F_{,b}+f_{,b}F_{,a}-f_{,c}F_{,c}\del_{ab})=0,
\nonumber\\
F_{,aa}+(4b+\f{e}{2}-1)\f{{F_{,a}}^2}{F}=0.
\label{2dgravred}
\end{eqnarray}
Identical conditions are obtained from other components in 
(\ref{eom} and (\ref{source}) as well. It can now be seen
that (\ref{2dgrav}) and (\ref{2dgravred}) are identical.
Indeed the LHS in equations (\ref{2dgrav}) or (\ref{2dgravred}) 
provide the explicit expressions for the terms that were 
dropped in writing $R_{ab}$ explicitly. We have therefore obtained
a necessary condition that our ansatz satisfies the supergravity 
equations of motion in eleven dimensions. 

Now, defining $\phi=\f{3e}{2}\ln F$, we can rewrite (\ref{2dgrav})
as:
\begin{eqnarray}
R_{ab}(\s{f}F^e)+\nabla_a\nabla_b\phi-\f{1}{3}\nabla_a\phi
\nabla_b\phi = 0,\nonumber\\
\nabla^2 \e^{-\phi} = 0.
\label{dilaton}
\end{eqnarray}
The above equations match with the equations obtained from the
following two dimensional dilaton gravity action\cite{cadoni}
provided $k=-\f{1}{2}$ and the cosmological term $\lambda$ 
is set to zero.
\begin{equation}
 S = -\int_{M} \s{g}\e^{-2\phi}[R+\f{8k}{k-1}
(\nabla\phi)^2+\lambda^2] -2\int_{\p M}\e^{-2\phi}K.
\label{dgaction}
\end{equation}
where $K$ is the trace of the second fundamental form, $\p M$  is the 
boundary of  $M$ and $k$ is a parameter taking values $|k|\leq 1$.
We therefore notice that a general class of membrane solutions can 
be constructed with our ansatz, for any solution of 2-dimensional 
gravity defined by equation (\ref{dgaction}). Since a world-volume 
dependent conformal factor $\s{f}$ appears explicitly in our solution,
the translational isometries in these directions, unlike the flat
membrane case, are now lost. In general we therefore have 
curved geometry for these branes.  

At this point, we make the remark that the relation between the membrane
source and the supergravity background turns out to be identical to that
of \cite{duff}. It will be useful to find the ADM mass and charge of
the solutions. 


\section{Discussions and Conclusions}

To recapitulate our results, we found a class
of curved membrane solutions in the eleven dimensional supergravity.
The worldvolume Poincar\'{e} invariance is broken while the isotropy  
in the directions transverse to the membrane survives.
The metric and the three form for the general solution are
given by :
\begin{eqnarray}
ds^2 = (1+\f{K}{F^{-{\f{3}{2}}e}{\T r}^6})^{-\f{2}{3}}F^e
[-(dx^0)^2+\s{f}(dx^a)^2]+(1+\f{K}{F^{-{\f{3}{2}}e}{\T
r}^6})^{\f{1}{3}}F^{-\f{e}{2}}(dy^{m})^2, \nonumber\\
A_{\mu \nu \rho} = \pm \f{1}{^{3}g}\varep_{\mu \nu \rho}
(1+\f{K}{F^{-{\f{3}{2}}e}{\T r}^6})^{-1}F^{{\f{3}{2}}e}\s{f}.
\label{oursoln}
\end{eqnarray}
This represents general curved membrane solution in a
class of embedding space. The geometry of the embedding
space can be obtained by taking the asymptotic limit
$r \rightarrow \infty$. 
We are motivated by the current 
upsurge of studies of 
nonsupersymmetric brane solutions 
in various contexts and are mainly 
interested in non-supersymmetric solutions. 
However, our solution space turns out to be 
large enough to contain several supersymmetric 
solutions as well. 

The
solution presented here can be used to obtain the curved membrane solution
in a class of embedding space  provided 
it satisfies some necessary condition for being a
solution which is the constraint imposed by (\ref{2dgravred}).
We can obtain the solution for various special cases,
including the curved membrane in the flat space
which is a special case of our general solution.
At this point it may be useful to compare our result
to that of \cite{papa}. They essentially use the conformal sigma
models and obtain curved transverse and longitudinal
geometries for higher and lower dimensional brane
respectively. Our approach is more straightforward 
and we consider only membrane with a curved longitudinal
world volume only. This method can be generalized for other
branes and can be used for explicit solutions of branes
wrapping cycles in asymptotically non-flat geometry.

One useful application of the present solution is
to consider the near horizon limit. 
For the flat brane by setting 
$f$ to unity and using the relation among the various exponents, 
we are left with essentially only one equation:
\begin{equation}
F_{,aa}-(\f{3e}{2}+1)\f{{F_{,a}}^2}{F}=0.
\label{2dfinal}
\end{equation}
which is the condition on the embedding space.
\footnote{Since the above equation(\ref{2dfinal}) is integrable, we can 
consistently set $f$ to unity.} 
Then we find that in the limit 
$K \rightarrow \infty$, the above metric goes over to 
the familiar $AdS_4 
\times S^7$. As the gravity dual depends only on the brane and
not on the embedding geometry, in our case we find that 
although the generic solution is non supersymmetric, we end up
with the same supergravity background as that of a supersymmetic 
solution. This analysis can be extended to generically  curved branes. 
The gravity duals in the generic case 
correspond to non conformal world volume theory and 
it is interesting to construct the dual of a confining theory.


Finally it is important to check the stability of these solutions.
Since these
solutions are not supersymmetric in general, their stability is not 
guaranteed. As a result, one would like to know if these solutions 
are stable against small perturbations of the metric. 
This is not so
innocuous a question as it might appear.
The gravitational 
stability
of black holes was a long standing problem\cite{regge}. In\cite{vishu},
it was settled in the affirmative for Schwarzschild black holes. 
Subsequently, the stability of Reissner-Nordstr\"{o}m\footnote
{A different type of instability still persists in the vicinity of the 
inner Cauchy horizon.}and Kerr black holes was also established. In low
energy string theory and M-theory, there is a plethora of black holes 
and p-branes. In particular, extended objects having event  horizons 
enclosing a curvature singularity emerge as classical solutions\cite{andy}
of supergravity. They can be thought of as higher dimensional cousins 
of the familiar black holes. It is thus of paramount importance to
study the stability of these solutions. In a series of papers\cite{GL}, 
it was argued that both uncharged as well as charged non-extremal black 
p-branes are unstable while the charged extremal black p-branes are stable.
As a specific example, a black string solution was considered and argued
to be unstable against decay into black holes. Recently, in \cite{gary},
the authors objected to this argument and proposed that such unstable
black string will finally settle down to some other black string. This 
proposal has the virtue that it is able to avoid the bifurcation of
event horizon. But there are still many unresolved questions. 
A full fledged perturbation analysis of the 
solutions presented in our paper is therefore important. 

We now end with some speculations.We presented the 
supergravity solutions which essentially provide the long distance 
bulk behavior. A complementary study of the short distance behavior
using matrix model \cite{bfss}, may be illuminating.
We also think that the Euclidean version of these branes might have some 
relevance to the brane world scenario. 
Finally, the similarity between the 
bosonic sector of the supersymmetric M-Theory and the Bosonic M-Theory in 
\cite{susskind}and the existence of nonsupersymmetric 2-branes in the Bosonic 
M-Theory makes us strongly feel that it might be possible to relate the 2-
brane solutions in these two apparently disparate theories. We hope to 
address some of these issues in future.
 
\section{Acknowledgements}
The work of S.M is supported by the National Science Foundation
under grant no. 9801875.

\section{Appendix}
For convenience, we provide the  Christoffel connections below
for our ansatz given in sections 2 and 3.
\begin{eqnarray}
& &{\G^0}_{0a} = -\f{1}{3}(\e^{3\T A}\p_a \e^{-3\T A}
-\f{3e}{2}\p_a \ln F), \nonumber\\
& &{\G^0}_{0m} = -\f{1}{3}\e^{3\T A}\p_m \e^{-3\T A},\nonumber\\
& &{\G^a}_{00} = -\f{1}{3}f^{-\f{1}{2}}(\e^{3\T A}\p_a \e^{-3\T A}
-\f{3e}{2}\p_a \ln F),\nonumber\\
& &{\G^a}_{bc} =-\f{1}{3}\e^{3\T A}F^{\f{3e}{2}}
[\p_c(\e^{-3\T A}F^{-\f{3e}{2}}){\del^a}_b+
\p_b(\e^{-3\T A}F^{-\f{3e}{2}}){\del^a}_c-\nonumber\\
& &\p_a(\e^{-3\T A}F^{-\f{3e}{2}})\del_{bc}]+\nonumber\\
& &\f{1}{4f}(\p_c f{\del^a}_b + \p_b f{\del^a}_c
- \p_a f \del_{bc}),\nonumber\\
& &{\G^a}_{bm} = -\f{1}{3}{\del^a}_b\e^{3\T A}\p_m\e^{-3\T A},
\nonumber\\
& &{\G^a}_{mn} = -\f{1}{2}(F^{-e}f^{-\f{1}{2}})\e^{-2\T A}
\p_a(\e^{2\T B}g^b)\del_{mn},\nonumber\\
& &{\G^m}_{00} =\f{1}{2}F^{e}g^{-b}\e^{-2\T B}\p_m\e^{2\T A},\nonumber\\
& &{\G^m}_{ab} = -\f{1}{2}(g^{-b}F^{e}f^{\f{1}{2}})\e^{-2\T B}
\p_m\e^{2\T A}\del_{ab},\nonumber\\
& &{\G^m}_{an} = \f{1}{2}\e^{-2\T B}g^{-b}\p_a(\e^{2\T B}g^b)
{\del^m}_n,\nonumber\\
& &{\G^l}_{mn} = \f{1}{6}\e^{-6\T B}[\p_n \e^{6\T B}
{\del^l}_m +\p_m \e^{6\T B}{\del^l}_n -\p_l \e^{6\T B}
\del_{mn}]
\label{chriscon}
\end{eqnarray}

The simplified form of these Christoffel connections(in the notation
explained in the text) are also given below.
\begin{eqnarray}
& &{\G^0}_{0a} = \f{d}{3}\f{K}{r^6}(1+\f{K}{r^6})^{-1}\f{h_{,a}}{h}+
\f{e}{2}\f{F_{,a}}{F}\nonumber\\
& &{\G^0}_{0m} = -\f{1}{3}(1+\f{K}{r^6})^{-1}(\f{K}{r^6})_{,m}\nonumber\\
& &{\G^a}_{00} =\f{d}{3}(\f{K}{r^6})(1+\f{K}{r^6})^{-1}\f{1}{\s f}
\f{h_{,a}}{h}+\f{e}{2\s f}\f{F_{,a}}{F}\nonumber\\
& &{\G^a}_{bc} =\f{d}{3h}\f{K}{r^6}(1+\f{K}{r^6})^{-1}(h_{,c}{\del^a}_b+
h_{,b}{\del^a}_c-h_{,a}\del_{bc})+\nonumber\\
& &\f{e}{2F}(F_{,c}{\del^a}_b+F_{,b}{\del^a}_c-F_{,a}\del_{bc})+\nonumber\\
& &\f{1}{4f}(f_{,c}{\del^a}_b+f_{,b}{\del^a}_c-f_{,a}\del_{bc})\nonumber\\
& &{\G^a}_{bm} = -\f{1}{3}(1+\f{K}{r^6})^{-1}(\f{K}{r^6})_{,m}{\del^a}_b
\nonumber\\
& &{\G^a}_{mn} = -\f{1}{2}(F^{-e}g^{b}f^{-\f{1}{2}})[-\f{d}{3}\f{K}{r^6}
\f{h_{,a}}{h}+b(1+\f{K}{r^6})\f{g_{,a}}{g}]\del_{mn}\nonumber\\
& &{\G^m}_{00} = -\f{1}{3}(F^{e}g^{-b})(1+\f{K}{r^6})^{-2}(\f{K}{r^6})_{,m}
\nonumber\\
& &{\G^m}_{ab} =\f{1}{3}(F^{e}g^{-b}\s f)(1+\f{K}{r^6})^{-2}
(\f{K}{r^6})_{,m}\del_{ab}\nonumber\\
& &{\G^m}_{an}= -\f{d}{6}\f{K}{r^6}(1+\f{K}{r^6})^{-1}\f{h_{,a}}{h}{\del^m}_n
+\f{b}{2}\f{g_{,a}}{g}{\del^m}_n\nonumber\\
& &{\G^m}_{np}= \f{1}{6}(1+\f{K}{r^6})^{-1}[(\f{K}{r^6})_{,p}{\del^m}_n+
(\f{K}{r^6})_{,n}{\del^m}_p - (\f{K}{r^6})_{,m}\del_{np}]
\label{mchriscon}
\end{eqnarray}
To present the relevant spin connections, let us first fix our notation. 
We denote the spacetime indices by $0$, $a$ and $m$ and the tangent
space indices by $\hat{0}$, $\hat{a}$ and $\hat{m}$. 
The nonvanishing spin connections are as follows.
\begin{eqnarray}
& &(\om_0)_{\hat{0}\hat{a}}= \f{1}{3}f^{-\f{1}{4}}(\e^{3\T A}
\p_{\hat{a}}\e^{-3\T A}-\f{3e}{2}\p_{\hat{a}} \ln F),\nonumber\\
& &(\om_0)_{\hat{0}\hat{m}}= -F^{\f{e}{2}}g^{-\f{b}{2}}\e^{-\T B}
\p_{\hat{m}}\e^{\T A},\nonumber\\
& &(\om_a)_{\hat{b}\hat{c}}= 
[\f{1}{3}\e^{3\T A}F^{\f{3e}{2}}
[\p_{\hat{b}}(\e^{-3\T A}F^{-\f{3e}{2}})\del_{a\hat{c}}-\nonumber\\
& &\p_{\hat{c}}(\e^{-3\T A}F^{-\f{3e}{2}})\del_{a\hat{b}}]-
\f{1}{4f}(\p_{\hat{b}}f \del_{a\hat{c}}-\p_{\hat{c}}f\del_{a\hat{b}})],
\nonumber\\
& &(\om_a)_{\hat{b}\hat{m}}=\f{1}{2}(F^{\f{e}{2}}f^{\f{1}{4}}
g^{-\f{b}{2}})\e^{-(\T A + \T B)}\p_{\hat{m}}\e^{2\T A},\nonumber\\
& &(\om_m)_{\hat{a}\hat{m}}= -\f{1}{2}(F^{-\f{e}{2}}f^{-\f{1}{4}}
g^{\f{b}{2}})\e^{-(\T A - \T B)}(\e^{-2\T B}\p_{\hat{a}}\e^{2\T B}+
g^{-b}\p_{\hat{a}} g^{b}),\nonumber\\
& &(\om_m)_{\hat{n}\hat{p}}= \e^{-\T B}(\p_{\hat{p}}\e^{\T B}
\del_{m\hat{n}}-\p_{\hat{n}}\e^{\T B}\del_{\hat{p}m}).
\label{spincon}
\end{eqnarray}

The above spin connections become simplified once we use the explicit
form of $\T A$, $\T B$ and $\T C$. 
\begin{eqnarray}
& &(\om_0)^{\hat{0}\hat{a}}= \f{d}{3}f^{-\f{1}{4}}\f{h_{,\hat a}}{h}
(1+\f{K}{r^6})^{-1},\nonumber\\
& &(\om_0)^{\hat{0}\hat{m}}=\f{1}{3}h^{-\f{3d}{2}}(1+\f{K}{r^6})^{-\f{3}{2}}
(\f{K}{{\T r}^6})_{,m},\nonumber\\
& &(\om_a)^{\hat{b}\hat{c}}= [-\f{d}{3}(1+\f{K}{r^6})^{-1}\f{1}{h}
(h_{,\hat c}\del_{a\hat b} - h_{,\hat b}\del_{a\hat c})+\nonumber\\
& &{\f{1}{4f}}(f_{,\hat c}\del_{a\hat b} - f_{,\hat b}\del_{a\hat c})],
\nonumber\\
& &(\om_a)^{\hat{b}\hat{m}}= -\f{1}{3}h^{-\f{3d}{2}}(1+\f{K}{r^6})^{-\f{3}{2}}
f^{-\f{1}{4}}{\del^{\hat b}}_a(\f{K}{{\T r}^6})_{,m},\nonumber\\
& &(\om_m)^{\hat{a}\hat{n}}= -\f{d}{6}h^{\f{d}{2}}f^{-\f{1}{4}}
(1+\f{K}{r^6})^{-\f{1}{2}}\f{h_{,\hat a}}{h}\del^{m\hat n},\nonumber\\
& &(\om_m)^{\hat{n}\hat{p}}= \f{1}{6}(1+\f{K}{r^6})^{-1}
[(\f{K}{r^6})_{,\hat p}\del_{m\hat n}-(\f{K}{r^6})_{,\hat n}\del_{m\hat p}].
\label{mspincon}
\end{eqnarray}

  

\begin{thebibliography}{99}
\bibitem{papa}
D.Brecher and M.J.Perry, ``Ricci-Flat Branes'',
Nucl.Phys.{\bf B566} (2000) 151-172, [hep-th/9908018];\\
B.Janssen, ``Curved branes and cosmological (a,b)-models'',
JHEP{\bf 0001} (2000) 044, [hep-th/9910077];\\
JM Figueroa-O'Farrill, ``More Ricci-flat branes'',
Phys.Lett.{\bf B471} (1999) 128-132, [hep-th/9910086];\\
G.Papadopoulos, J.G.Russo and A.A.Tseytlin, 
``Curved Branes from String Dualities'',
Class.Quant.Grav. 17 (2000) 1713-1728, [hep-th/9911253].
\bibitem{kabat}
 A.L.Larsen and C.O.Lousto, ``On the stability of
spherical membranes in curved spacetimes'', Nucl.Phys.{\bf B472}
(1996) 361-376, [gr-qc/9602009];\\
D.Kabat and W.Taylor, ``Spherical membranes in Matrix
theory'',\\ Adv.Theor.Math.Phys.{\bf 2} (1998) 181-206,
[hep-th/9711078].
\bibitem{douglas}M.Douglas, ``D-branes in Curved Space'', hep-th/9703056.
\bibitem{wzw}A.Y.Alekseev and V.Schomerus, ``D-branes in the WZW model'',
Phys.Rev.{\bf D60} (1999) 061901, [hep-th/9812193] and references 
therein.  
\bibitem{douglas-bachas}C.Bachas, M.Douglas and C.Schweigert, 
``Flux Stabilization of D-branes'', JHEP {\bf 0005} (2000) 048, 
[hep-th/0003037].
\bibitem{julia}E.Cremmer, B.Julia and J.Scherk, ``Supergravity Theory
in eleven dimensions'', Phys.Lett.{\bf B76} (1978) 409.
\bibitem{sezgin}E.Bergshoeff, E.Sezgin and P.K.Townsend,
``Supermembranes and eleven\\ dimensional Supergravity'', 
Phys.Lett.{\bf B189} (1987) 75.
\bibitem{duff}M.J.Duff and K.S.Stelle, ``Multimembrane Solutions of
D=11 Supergravity'', Phys.Lett.{\bf B253} (1991) 113.
\bibitem{witten}E.Witten, ``String Theory Dynamics In Various Dimensions'', 
Nucl.Phys. {\bf B443} (1995) 85, [hep-th/9503124].
\bibitem{dabholkar}A.Dabholkar, G.Gibbons, J.A.Harvey, F.Ruiz-Ruiz,
``Superstrings and Solitons'', Nucl.Phys.{\bf B340} (1990) 33.
\bibitem{cadoni}M.Cadoni and S.Mignemi, ``Dilatonic black holes in
theories with moduli fields'', Phys.Rev.{\bf D48} (1993) 5536,
[hep-th/9305107]; A.Kumar and K.Ray, ``Thermodynamics of Two
Dimensional Black Holes'', Phys.Lett.{\bf B351} (1995) 431,
[hep-th/9410068].
\bibitem{susskind}G.T.Horowitz and L.Susskind, ``Bosonic M Theory'',
J.Math.Phys. {\bf 42} (2001) 3152-3160,[hep-th/0012037]. 
\bibitem{regge}T.Regge and J.A.Wheeler, ``Stability of a Schwarzschild 
singularity'',\\ 
Phys.Rev.{\bf 108} (1957) 1063.
\bibitem{vishu}C.V.Vishveshwara, ``Stability of the Schwarzschild Metric'', 
Phys.Rev.{\bf D1} (1970) 2870.
\bibitem{andy}G.T.Horowitz and A.Strominger, ``Black Strings and p-Branes'',\\
Nucl.Phys.{\bf B360} (1991) 197.
\bibitem{GL}R.Gregory and R.Laflamme, ``Black Strings and p-Branes are 
Unstable'',\\ Phys.Rev.Lett{\bf 70} (1993) 2837, [hep-th/9301052];\\  
``The Instability of Charged Black Strings and
p-Branes'',\\ Nucl.Phys.{\bf B428} (1994) 399, [hep-th/9404071];\\
``Evidence for the Stability of Extremal Black p-Branes'',\\ 
Phys.Rev.{\bf D51} (1995) 7007, [hep-th/9410050].
\bibitem{gary}G.T.Horowitz and K.Maeda, ``Fate of the Black String
Instability'',\\ Phys.Rev.Lett{\bf 87} (2001) 131301, [hep-th/0105111].
\bibitem{bfss} T.Banks, W.Fischler, S.Shenker and L.Susskind, 
``M-theory as a matrix model: a conjecture'',
Phys.Rev.{\bf D55} (1997) 5112, [hep-th/9610043].
\end{thebibliography}
\end{document}